\documentclass[a4paper,11pt]{article}   	

\usepackage{jheppub} 

\usepackage[T1]{fontenc} 

\usepackage{extarrows}
\usepackage{mathtools}

\usepackage{amsmath,amsfonts,amssymb,epsfig}
\usepackage{slashed}

\usepackage{bbold}

\usepackage{cancel}
\usepackage{bm}
\usepackage{multirow}

\newcommand{\exclude}[1]{}
\title{\boldmath  Salvage of Too Slow Gravitinos}

\author[a,b,1]{ I. Antoniadis,}
\author[a,2]{K. Benakli,}
\author[a,3]{and W. Ke}

\affiliation[a]{ Laboratoire de Physique Th\'eorique et Hautes Energies (LPTHE),\\ UMR 7589,
Sorbonne Universit\'e et CNRS, \\ 4 place Jussieu, 75252 Paris Cedex 05, France}
\affiliation[b]{Institute for Theoretical Physics, KU Leuven, Celestijnenlaan 200D, B-3001 Leuven, Belgium}

\emailAdd{antoniad@lpthe.jussieu.fr}
\emailAdd{kbenakli@lpthe.jussieu.fr}
\emailAdd{wke@lpthe.jussieu.fr}

\abstract{Gravitinos can inherit a non-relativistic dispersion relation while propagating in a background breaking both supersymmetry and Lorentz symmetry spontaneously. This is because the longitudinal mode velocity is controlled by the sound speed in the background. It was pointed out recently by Kolb, Long and McDonough that the production of gravitinos might diverge when this sound speed vanishes. We argue that in the framework of  cosmological models with linearly spontaneously broken realised supersymmetry, where the physical fermions are combinations of the vacuum goldstino and the inflatino, the gravitino longitudinal mode has a relativistic dispersion relation and therefore avoids the catastrophic production. We illustrate this in some explicit examples.}

\begin{document} 
\maketitle
\flushbottom

\section{Introduction}
\label{Sec:introduction}

In the super-Higgs mechanism,  a  spin $1/2$ fermion, the goldstino, combines with the gravitino and provides it with the appropriate number of degrees of freedom for a massive spin 3/2  \cite{Fayet:1974jb,Volkov:1973jd,Fayet:1977vd,Deser:1977uq,Cremmer:1978hn}.  When the spin 1/2  propagates in a generic background with a non-relativistic dispersion relation, for instance at the sound speed $c_s <1$ in fluids, the result of the super-Higgs mechanism has been denoted "slow gravitinos" \cite{Benakli:2014bpa,Benakli:2013ava,Benakli:2015mbb,Kahn:2015mla,Ferrara:2015tyn}. Such situations might occur in cosmological backgrounds as cosmological solutions treat time and space differently \cite{Kallosh:1999jj,Giudice:1999yt,Giudice:1999am,Schenkel:2011nv,Kallosh:2000ve}.

At the end of inflation, during the period of reheating, the inflaton dissipates its energy while oscillating around the minimum of its potential. This energy is in part converted into a non-thermal production of gravitinos. This process was studied in \cite{Kallosh:1999jj,Giudice:1999yt,Giudice:1999am,Kallosh:2000ve} where the equations of propagation of the different gravitino modes were established. In particular, there is a copious production of the helicities $\pm 1/2$ components of the gravitinos, the goldstino fermion at the moment, which was also numerically evaluated in those papers and in \cite{Nilles:2001fg,Nilles:2001ry}. 

Recently, \cite{Kolb:2021xfn,Kolb:2021nob} have reconsidered this process. The equation of motion of the fermion $\theta$ describing the longitudinal mode in the vaccuum after reheating can be written as
\begin{equation}
\left[\bar{\gamma}^{0} \partial_{0}+\mathrm{i}\bar{\gamma}^{i} k_{i} c_s \right] 
\theta + \cdots =0,
\label{vs of theta}
\end{equation}
where the $\cdots$ stand for mass and mixing terms with other fermions. The sound speed is identified then as given by:
\begin{equation}
c_{s}^{2}=\frac{\left(p-3 m_{3/2}^{2} M_{P}^{2}\right)^{2}}{\left(\rho+3 m_{3/2}^{2} M_{P}^{2}\right)^{2}}+\frac{4M_{P}^{4}\left(  \partial  m_{3/2}/\partial t\right)^{2}}{\left(\rho+3 m_{3/2}^{2} M_{P}^{2}\right)^{2}}\,,\label{Sound-speed-1}
\end{equation}
where $M_P$ is the reduced Planck mass, $m_{3/2}$ the gravitino mass, while $\rho,p$ denote the energy density and pressure of the matter system.
It was then noticed in \cite{Kolb:2021xfn,Kolb:2021nob}  that a catastrophic gravitino production occurs whenever the above sound speed vanishes. 

In the supergravity cosmological models we discuss here where supersymmetry is linearly realised (but spontaneously broken), the slow goldstino is not an eigenstate of the Hamiltonian. It mixes with the fermion whose scalar partner has a non-vanishing kinetic energy, for instance the inflatino in the works mentioned above. These give rise at each moment, after diagonalisation, to combinations of the two fermions that are eigenstates of the Hamiltonian and $c_s$ is not a physical quantity. This has motivated \cite{Ferrara:2015tyn} to resort to non-linear realisation of supersymmetry in order to project out the second fermion and construct models of slow gravitinos. If one keeps both states, it was pointed out in \cite{Dudas:2021njv} that $c_s^2=0$ does not lead to catastrophic gravitino production. More precisely, as the coefficients of the momentum vector $\vec{k}$ in the kinetic term are now described by a mixing matrix that is known not to be singular as $c_s^2=0$, no catastrophic production is expected. 

The aim of this work is to derive explicitly the general dispersion relations of the physical fermions in standard linear supergravity and show that not only as expected the mixing matrix in the fermion kinetic terms is not singular, but, it can be diagonalised to the identity matrix. Using the non-adiabaticity coefficient, we conclude that in this background catastrophic production does not occur. Furthermore, we show in particular cases that the physical sound speed is equal to one independently of the value of $c^2_s$; namely, for degenerate fermion masses, the dispersion relation is relativistic despite the time dependent gravitational  background.

In section 2, we consider the general case with a goldstino combination of two fermions. We exhibit through diagonalisation of the kinetic terms the generic form of the dispersion relations of the physical fermions. We argue that these do not lead to catastrophic gravitino production.  One of the novelties of our work is to consider also the case where one of the fermions arises from a vector multiplet. Section 3 presents some examples where the whole diagonalisation can be carried out explicitly and where the dispersion relation takes a relativistic form.

\section{The General case with two fermions}
\label{sec:diago}
We consider an $N=1$ supergravity model where in addition to the graviton and gravitino $\psi_{\nu}$, one has two possible sources of supersymmetry breaking {\it in the vacuum}. The first is a potentially non-vanishing $D$-term $\mathcal{P}$ for a $U(1)$ vector multiplet. The second possibility uses the non-vanishing $F$-term of a chiral multiplet. In addition, during the cosmological evolution, there is an extra source of supersymmetry breaking given by the non-vanishing kinetic energy of a rolling scalar, the inflaton. To describe this system we consider a vector supermultiplet with field strength and gaugino denoted as $F_{\mu \nu}$ and $\lambda$, respectively, as well as  one or two chiral multiplets consisting in scalars $\phi_i$ and fermions $\chi_i$, $i=1,2$. The corresponding Lagrangian is given by: \footnote{The Lagrangian with a generic number of chiral and vector multiplets can be found in \cite{Kallosh:2000ve}.}
\begin{equation}
\begin{aligned}
e^{-1} \mathcal{L} &=-\frac{1}{2} M_{P}^{2} R-g_{i}^{j}\left(\hat{\partial}_{\mu} \phi^{i}\right)\left(\hat{\partial}^{\mu} \phi_{j}\right)-V \\
&-\frac{1}{2} M_{P}^{2} \bar{\psi}_{\mu} R^{\mu}+\frac{1}{2} m \bar{\psi}_{\mu R} \gamma^{\mu \nu} \psi_{\nu R}+\frac{1}{2} m^{*} \bar{\psi}_{\mu L} \gamma^{\mu \nu} \psi_{\nu L}\\
&+\left(\operatorname{Re} f \right)\left[-\frac{1}{4} F_{\mu \nu} F^{\mu \nu }-\frac{1}{2} \bar{\lambda} \mathcal{D} \lambda\right]
+\frac{1}{4} \mathrm{i}\left(\operatorname{Im} f \right)\left[F_{\mu \nu} \tilde{F}^{\mu \nu }-\hat{\partial}_{\mu}\left(\bar{\lambda} \gamma_{5} \gamma^{\mu} \lambda\right)\right] 
 \\
&+\frac{1}{4}\left\{\left(\operatorname{Re} f\right) \bar{\psi}_{\mu} \gamma^{\nu \rho} F_{\nu \rho} \gamma^{\mu} \lambda-\left[f^{i} \bar{\chi}_{i} \gamma^{\mu \nu} F_{\mu \nu}^{-} \lambda_{L}+\mathrm{h.c.}\right]\right \} \\
&-g_{i}{}^{j}\left[\bar{\chi}_{j} \mathcal{D} \chi^{i}+\bar{\chi}^{i} \mathcal{D} \chi_{j}\right]-m^{i j} \bar{\chi}_{i} \chi_{j}-m_{i j} \bar{\chi}^{i} \chi^{j}\\
&-2 m_{i\alpha} \bar{\chi}^{i} \lambda-2 m^{i\alpha} \bar{\chi}_{i} \lambda-m_{R, \alpha \beta} \bar{\lambda}_{R} \lambda_{R}-m_{L, \alpha \beta} \bar{\lambda}_{L} \lambda_{L} \\&+\left(2 g_{j}{}^{i} \bar{\psi}_{\mu R} \gamma^{\nu \mu} \chi^{j} \hat{\partial}_{\nu} \phi_{i}+\bar{\psi}_{R} \cdot \gamma \upsilon_{L}+\mathrm{h.c.}\right)
\end{aligned}\label{lagrangian}
\end{equation}
where $L,R$ subscripts refer to the left and right chiralities, respectively. Moreover, $\chi_i$ is a left-handed field while $\chi^i$ is right-handed, and $\phi^i$ denotes the complex conjugate of $\phi_i$. The kinetic term of the gravitino is defined as $R^{\mu}=\gamma^{\mu \rho \sigma} \mathcal{D}_{\rho} \psi_{\sigma}$. The covariant derivatives as well as the mass terms in this Lagrangian can be found in Appendix \ref{appendix}. The Greek index $\alpha$ in the gaugino mass terms $m_{i\alpha}$ is set to 1, since there is at most one vector multiplet, in which case there is also only one chiral multiplet and thus the latin index $i$ is also set to 1. The field strengths $F_{\mu\nu},\tilde{F}^{\mu\nu}, F_{\mu\nu}^-$ are irrelevant for our discussion, and are defined in \cite{Kallosh:2000ve}.

The K\"ahler metric $g_i{}^j$ is given by the K\"ahler potential $K$ with
\begin{equation}
    g_i{}^j=\frac{\partial}{\partial \phi^{i}} \frac{\partial}{\partial \phi_{j}} K
\end{equation}
and the gravitino mass $m_{3/2}$ is determined by the superpotential $W$ as well as the K\"ahler potential
\begin{equation}
    m_{3/2}=|m| M_{\mathrm{P}}^{-2} ,\quad m \equiv \mathrm{e}^{\frac{K}{2 M_{\mathrm{P}}^{2}}} W
\end{equation}
The scalar potential is a sum of the $F$-term and $D$-term contributions, with $m_i$ the K\"ahler covariant derivative of $m$ and $f \equiv 1/g^2$ denoting the gauge kinetic function, assumed to be constant:
\begin{equation}
    V=V_F+V_D,\quad V_F=-3 M_{\mathrm{P}}^{-2}|m|^{2}+m_{i} \left(g_{j}{}^i\right)^{-1} m^{j},\quad V_D=\frac{1}{2}g^2\mathcal{P}^2
\label{potential}
\end{equation}
where $\mathcal{P}$ is the Killing potential.

In the following, we will consider a flat universe described by the Friedmann-Lema\^itre-Robertson-Walker (FLRW) metric $\mathrm{d} s^{2}=a^{2}(\eta)\left(-\mathrm{d} \eta^{2}+ \mathrm{d} \mathbf{x}^{2}\right)$ where $a$ is the scale factor and  $\eta\equiv x^0$ is the conformal time. The determinant of the vierbein is then $e=a^4$. We further introduce the dot derivative with respect to the physical time $t$, with $\dot{\mathrm{f}}\equiv  a^{-1}\partial_0 \mathrm{f}$. The Hubble rate is defined as $H\equiv \dot{a}/a$. 

Throughout this work, we will assume real backgrounds, and use the plane wave expansion for the fermions $\Psi(\eta,\vec{k})=\exp (\mathrm{i}\vec{k}\cdot\vec{x})\Psi(\eta)$. Useful notations are:
\begin{equation}
    \begin{aligned}&\alpha\equiv \rho+3 M_{P}^{-2}m^{2},\quad 
   \alpha_{1} \equiv p-3 M_{P}^{-2}m^{2},\quad \alpha_{2} \equiv 2 \dot{{m}},\\&\hat{A}=\hat{A}_1+\bar{\gamma}^0 \hat{A}_2\equiv \frac{1}{\alpha}\left(\alpha_{1}+\bar{\gamma}^{0} \alpha_{2}\right),\\& \hat{B}=\hat{B}_1+\bar{\gamma}^0 \hat{B}_2\equiv -\frac{3}{2} \dot{a} \hat{A}-\frac{1}{2} M_{P}^{-2} {m} a \bar{\gamma}^{0}(1+3 \hat{A})\\&n_{i}\equiv g_{i}{}^{j} \dot{\phi}_{j}, \quad n^{i}\equiv g_{j}{}^{i} \dot{\phi}^{j} ,\quad |\dot{\phi}|^{2} \equiv g_{i}{}^{j} \dot{\phi}_{j} \dot{\phi}^{i}\\&\xi^{i} \equiv m^{i}+\bar{\gamma}^{0} n^{i},\quad \xi_{i} \equiv m_{i}+\bar{\gamma}^{0} n_{i}\\&\Delta^2 \equiv 1-\frac{\alpha_1^2}{\alpha^2}-\frac{\alpha_2^2}{\alpha^2}
=\frac{4}{\alpha^2} \left[\dot{\phi}^{i} \dot{\phi}_{j} m_{k} m^{\ell}\left({g_\ell{}^{ k}}^{-1} g_{i}{}^{j}-\delta_{i}^{k} \delta_{\ell}^{j}\right)+\frac{1}{2}|\dot{\phi}|^{2}g^2 \mathcal{P}^2 \right]  \end{aligned}
\label{notations}
\end{equation}
In the FLRW background, the energy density and pressure are given in terms of the Hubble parameter as
\begin{equation}
    \rho=3 M_{P}^{2} H^{2}, \quad p=-M_{P}^{2}\left(3 H^{2}+2 \dot{H}\right)
\end{equation}

Before choosing a gauge, the goldstino $\upsilon$ in the last line of \eqref{lagrangian} takes the form
\begin{equation}
 \upsilon=\xi^{\dagger i} \chi_{i}+\xi_{i}^{\dagger} \chi^{i}+\frac{\mathrm{i}}{2}  \gamma_{5} \mathcal{P} \lambda\label{goldstino}
\end{equation}
To describe the theory in the supersymmetry broken phase, we follow \cite{Kallosh:2000ve} and introduce the combination of spin-$1/2$ fermions 
\begin{equation}
\begin{aligned}
& \theta \equiv \bar{\gamma}^i {\psi}_i\quad ,\quad \Upsilon \equiv a\left(n_{i} \chi^{i}+n^{i} \chi_{i}\right)
\\&\Xi_{R}=-m^{k} g_{k}^{-1 j} m_{j i} \chi^{i}+\bar{\gamma}^{0} \dot{\phi}_{j}\left(m^{j i} \chi_{i}+m^{j} \lambda_{L} \right)+\mathrm{i} M_{P}^{-2} m \mathcal{P} \lambda_{R}-\mathrm{i}g^2 \mathcal{P} m_{i} \chi^{i}
\end{aligned}\label{3fermions}
\end{equation}

In the unitary gauge $\upsilon=0$, \eqref{goldstino} then relates the gaugino to the chiral fermions. The spinors in \eqref{3fermions} are \textit{a priori} independent. However, as we consider the case of two fermions, $\theta$ will be associated with the longitudinal component in the vaccum of the gravitino in the unitary gauge. The fermion $\Upsilon$ describes the correction to this mode from supersymmetry breaking by the rolling scalar kinetic energy. In this case, $\Upsilon$ and $\Xi$ are proportional to each other, with 
\begin{equation}
    \Xi=-a^{-1} \hat{F} \Upsilon\label{def:F}
\end{equation}
For two chiral multiplets, the matrix $\hat{F}$ is provided in \cite{Kallosh:2000ve}. In the presence of a $D$-term, the remaining chiral multiplet is written as $(\chi_1, \phi_1)$, and the kinetic energy becomes $|\dot{\phi}|^2=g_1{}^1\dot{\phi}_1^2$ . We find (for non-vanishing $|\dot{\phi}|^2$ and $\mathcal{P}$):
\begin{equation}
    \begin{aligned}
    \hat{F}=& \frac{\dot{V}}{2|\dot{\phi}|^2}-\frac{\dot{\mathcal{P}}}{\mathcal{P}} + \bar{\gamma}^0\left((g_1{}^1)^{-1} m_{11} -  \frac{\dot{\mathcal{P}}}{\mathcal{P}}\frac{m^1}{n^1}+\frac{2m}{M_P^2} \right)
    \end{aligned}\label{F-Dterm}
\end{equation}

As emphasised by \cite{Kallosh:2000ve,Nilles:2001fg,Nilles:2001ry}, the equations of motions for $\theta$ and $\Upsilon$ are coupled together, thus  spin-$1/2$ particles produced are not necessarily the longitudinal component of the gravitino, but the fermions that diagonalise the Hamiltonian. We call them the \textit{physical fermions} thereafter.
Moreover, the spin-${1}/{2}$ fermions ($\theta$, $\Upsilon$) have non-canonical kinetic terms and thus will be rescaled before diagonalising their equations of motion. It was noticed in \cite{Nilles:2001fg,Nilles:2001ry}, in the two chiral multiplets case, that the kinetic terms can be
written as
\begin{equation}
    \mathcal{L}\supset  -\frac{4 a}{\alpha \Delta^{2}} \bar{\Upsilon} \bar{\gamma}^{0} \partial_{0} \Upsilon -\frac{\alpha}{4 k^{2}} a^{3} \bar{\theta} \bar{\gamma}^{0} \partial_{0} \theta\label{kinetic}
\end{equation}

Here we are going to generalise the above expression in the presence of a vector multiplet with a non-vanishing $D$-term. $\Upsilon$ can be projected onto the left-handed gaugino by
\begin{equation}
\begin{aligned}
P_{L} \xi^{\dagger 1} \Upsilon &=-\frac{1}{2} a P_{L} \xi^{\dagger 1} \bar{\gamma}^{0}\left(\xi_{1} \chi^{1}+\xi^{1} \chi_{1}+\frac{\mathrm{i}}{2} \gamma_{5} \mathcal{P} \lambda\right) \end{aligned}
\end{equation}
where in the first line we used the unitary gauge condition $\upsilon=0$ from \eqref{goldstino}. The right-handed gaugino is obtained by charge conjugation:
\begin{equation}
P_{R} \xi_{1}^{\dagger} \Upsilon=\frac{\mathrm{i}}{2} a n_{1} \mathcal{P} P_{R} \lambda
\end{equation}
On the other hand, it is easier to project $\Upsilon$ onto the chiral fermions given its definition:
\begin{equation}
\chi_{1}=\frac{P_{L} \Upsilon}{a n^{1}}\quad , \quad \chi^{1}=\frac{P_{R} \Upsilon}{a n_{1}}
\label{Up-kin}\end{equation}
Expressing the gaugino and chiral fermion kinetic terms in terms of $\Upsilon$, we find 
\begin{equation}
    \mathcal{L}\supset -\frac{4a V_D}{\alpha^2 \Delta^2}\left(\bar{\Upsilon} \bar{\gamma}^0\partial_0\Upsilon \right) -\frac{4a(\alpha-V_D)}{\alpha^2\Delta^2}\bar {\Upsilon}\bar{\gamma}^0\partial_0 \Upsilon= -\frac{4 a}{\alpha \Delta^{2}} \bar{\Upsilon} \bar{\gamma}^{0} \partial_{0} \Upsilon 
\label{Up-kin2}\end{equation}

As for $\theta$, one uses the fact that the spatial component of the gravitino can be decomposed into
\begin{equation}
\vec{\psi}=\vec{\psi}^{T}+\frac{1}{{k}^{2}}\left[\vec{k}(\bar{\gamma}^ik_i)+\frac{1}{2} \mathrm{i}(3 \vec{k}-\vec{\gamma}(\bar{\gamma}^ik_i))\left(\dot{a} \bar{\gamma}^{0}+M_{P}^{-2} a {m}\right)\right] \theta
\end{equation}
where $\psi^T$ corresponds to the transverse mode. Inserting the above equation into the gravitino kinetic term in the Lagrangian, we recover the same form as in  \eqref{kinetic}. Consequently, in presence of one chiral multiplet and one vector multiplet, the kinetic terms of $\theta$ and $\Upsilon$ are the same as for two chiral multiplets, up to a redefinition of $\alpha$ and $\Delta$. We thus can use the same rescaling in the two cases, allowing to have canonical fields $\{\Psi_1, \Psi_2\}$:
\begin{equation}
\theta=\frac{2 \mathrm{i} \bar{\gamma}^{i} k_{i}}{\left(\alpha a^{3}\right)^{1 / 2}} \Psi_1 ,\quad 
\Upsilon=\frac{\Delta}{2}\left(\frac{\alpha}{a}\right)^{1 / 2} \Psi_2
\end{equation}

\subsection{The mixing matrices}

In the basis $\{\Psi_1,\Psi_2\}$, the spin-$\frac{1}{2}$ part of the Lagrangian takes the form:
\begin{equation}
    \begin{aligned}
    \mathcal{L}_{\Psi_1 \Psi_2}=&-\bar{\Psi}_1 \left[ \bar{\gamma}^0\partial_0 \Psi_1 -\frac{1}{2}\frac{\partial_0(\alpha a^3)}{\alpha a^3}\bar{\gamma}^0\Psi_1 + \bar{\gamma}^0 \hat{B}\Psi_1 +\mathrm{i}\bar{\gamma}^i k_i\hat{A}^\dagger \Psi_1-i\bar{\gamma}^i k_i\Delta\bar{\gamma}^0 \Psi_2\right]\\
    &-\bar{\Psi}_2\left[ \bar{\gamma}^0\partial_0 \Psi_2 +\frac{\partial_0\left( \Delta \sqrt{\frac{\alpha}{a}}\right)}{\Delta \sqrt{\frac{\alpha}{a}}}\bar{\gamma}^0\Psi_2 + \bar{\gamma}^0 \hat{B}^\dagger \Psi_2 +\mathrm{i}\bar{\gamma}^i k_i\hat{A}\Psi_2+2\dot{a}\bar{\gamma}^0\Psi_2\right. \\&\quad \left.+\frac{am}{M_P^2}\Psi_2 + a\bar{\gamma}^0\hat{F}\Psi_2+\mathrm{i}\bar{\gamma}^0\bar{\gamma}^i k_i\Delta \Psi_1\right]
    \end{aligned}
\label{canoL}
\end{equation}
where the different parameters are given in \eqref{notations} and  $\hat{F}=\hat{F}_1+\bar{\gamma}^0\hat{F}_2$ is defined in \eqref{def:F}.  The explicit form of \eqref{canoL} depends on the specific model. We will illustrate some examples in Section \ref{sec:examples}.

With $\Psi$ designating the vector $(\Psi_1,\Psi_2)^T$, the above Lagrangian can be put in a simple form 
\begin{equation}
    \mathcal{L}_{\Psi_1 \Psi_2}=-\bar{\Psi}\left[\bar{\gamma}^{0}
    \partial_{0}+\mathrm{i}\bar{\gamma}^{i} k_{i}N +M\right] \Psi
\end{equation}
with equations of motion
\begin{equation}
\left[\bar{\gamma}^{0} \partial_{0}+\mathrm{i}\bar{\gamma}^{i} k_{i}N +M\right]_{mn} \Psi_n=0,\quad m,n\in\{1,2\}.
\end{equation}
The $N$ and $M$ mixing matrices are given by\footnote{The difference of some signs compared to \cite{Nilles:2001fg,Nilles:2001ry,Dudas:2021njv} is due to the $\bar{\gamma}^0$ convention.}
\begin{equation}\begin{aligned}M&=\mathbb{1}_4\left(
    \begin{array}{cc}
    -\hat{B}_2     & 0 \\
         0&\hat{B}_2+\frac{am}{M_P^2}-a\hat{F}_2\! \!
    \end{array}\right)+\bar{\gamma}^0\left( \begin{array}{cc}
       -\frac{1}{2}\frac{\partial_0(\alpha a^3)}{\alpha a^3}+\hat{B}_1  & 0 \\
         0&  \frac{\partial_0\left( \Delta \sqrt{\frac{\alpha}{a}}\right)}{\Delta \sqrt{\frac{\alpha}{a}}} + \hat{B}_1 +2\dot{a}  +a \hat{F}_1\! \!
    \end{array}\right)\\&=\mathbb{1}_4\left(
    \begin{array}{cc}
    -\hat{B}_2     & 0 \\
         0&\hat{B}_2+\frac{am}{M_P^2}-a\hat{F}_2 
    \end{array}\right)
\end{aligned}
\label{M-mat}
\end{equation}
and
\begin{equation}
    N=N_1+\bar{\gamma}^0 N_2=\mathbb{1}_4\left(\begin{array}{cc}
    \hat{A}_1     & 0 \\
        0 & \hat{A}_1
    \end{array} \right)+\bar{\gamma}^0 \left(\begin{array}{cc}
    -     \hat{A}_2 & -\Delta \\
         -\Delta & \hat{A}_2
    \end{array} \right)\label{N-mat}
\end{equation}
where $\mathbb{1}_4$ is a $4\times 4$ unit matrix. We should stress that in the decompositions of $M$ and $N$, the $2\times2$ matrices act on the basis $\{\Psi_1,\Psi_2\}$, while $\bar{\gamma}^0$ is a $4\times4$ matrix acting on the spinor indices.
In the first line of \eqref{M-mat}, the $\bar{\gamma}^0$-dependent part vanishes both for two chiral multiplets and for one chiral multiplet with a $D$-term due to the property 
\begin{equation}
    \hat{B}_1=\frac{1}{2}\left( \hat{B}+\hat{B}^\dagger\right)=\frac{a\dot{\alpha}}{2\alpha}+\frac{3\dot{a}}{2}
\end{equation}
Note also that for $\Delta$ appearing in the denominator of $\Upsilon$ kinetic term \eqref{Up-kin2}, there seems to be a singularity at $\Delta\rightarrow0$, but this singularity cancels out in the mixing matrices $M,N$, because there is another $\Delta$ in the denominator of $\hat{F}_1$ compensating the one in front of the $\Upsilon$ kinetic term.

In the $\{\Psi_1,\Psi_2\}$ basis, only $N$ contributes to the mixing. One might be tempted to diagonalize $N$ so as to decouple the two fermions, but in general the mixing matrices depend on time, thus a unitary transformation to the basis diagonalising $N$ would also be time-dependent, which by time derivative gives a contribution to the mass matrix, rendering $M$ non-diagonal.
Though on general grounds, we will not provide an analytical expression of the physical fermions in terms of $\{\Psi_1, \Psi_2\}$, as long as we consider the catastrophic production  in \cite{Kolb:2021nob,Kolb:2021xfn} it is not necessary to carry out the entire diagonalization of \eqref{canoL}.

\subsection{Dispersion relations for the physical fermions}\label{sec:dispersion}

Expressing \eqref{Sound-speed-1} in terms of the parameters of \eqref{notations}, we observe that the sound speed \eqref{Sound-speed-1} amounts to the norm of $N_{11}$, namely
\begin{equation}
c_s^2=\hat{A}_1^2+\hat{A}_2^2=1-\Delta^2    \label{kolb-speed}
\end{equation}
In the case of a single chiral multiplet, the fields $\Upsilon$ and $\Xi$ vanish in the unitary gauge, and we are left with $\theta$. The norm of $N\equiv N_{11}$ enters into the dispersion relation as the gravitino velocity. It is a well-known result \cite{Kallosh:1999jj,Giudice:1999yt,Giudice:1999am} that $\hat{A}_1^2+\hat{A}_2^2=1$ for one chiral multiplet, thus the gravitino sound speed is the speed of light.
However, when two fermions are present, $\Upsilon$ cannot be omitted and the physical fermions are combinations of $\theta$ and $\Upsilon$.  The question is then raised whether  \eqref{kolb-speed}  is the sound speed of a \textit{physical} propagating state.

One can check that the mixing matrix $N$  in \eqref{N-mat} is unitary, thus it can be written as an exponential of a phase
\begin{equation}
    N=\exp \left(2 \Phi \bar{\gamma}^{0}\right)=\cos(2\Phi)+\bar{\gamma}^0\sin(2\Phi), \quad \Phi^{\dagger}=\Phi
\label{N-exponential}
\end{equation}
Furthermore, notice that $N_1$ and $N_2$ are real, then $\Phi$ is a real, thus symmetric matrix. By a unitary transformation $\hat{\Psi}=\operatorname{exp}(\bar{\gamma}^0\Phi)\Psi$, the exponent in (\ref{N-exponential}) is taken away, making $N$ equal to the identity, and the Lagrangian \eqref{canoL} in the new basis $\hat{\Psi}=(\hat{\Psi}_1,\hat{\Psi}_2)^T$ becomes
\begin{equation}
    \mathcal{L}_{\hat{\Psi}_1\hat{\Psi}_2}=-\bar{\hat{\Psi}}\left[ \bar{\gamma}^0 \partial_0+ \mathrm{i}\bar{\gamma}^i k_i + \hat{M}\right]\hat{\Psi}\,.
\label{lagrangian-hat}
\end{equation}
The new mass matrix 
\begin{equation}
\begin{aligned}
&\hat{M}=\hat{M}_1+\bar{\gamma}^0\hat{M}_2=\operatorname{exp}(\bar{\gamma}^0\Phi)M \operatorname{exp}(-\bar{\gamma}^0\Phi)+\partial_0 \Phi    \\&\hat{M}_1=\operatorname{cos}(\Phi)M\operatorname{cos}(\Phi)+\operatorname{sin}(\Phi)M\operatorname{sin}(\Phi)+\partial_0 \Phi ,\\& \hat{M}_2=\operatorname{sin}(\Phi)M\operatorname{cos}(\Phi)-\operatorname{cos}(\Phi)M\operatorname{sin}(\Phi)
\end{aligned}\label{M-hat}\end{equation}
is in general non-diagonal, due to the off-diagonal elements of $\Phi$. We obtain therefore a system of two propagating fermions subject to oscillations due to the time-dependent mixing in their mass matrix. 

As a side remark, the matrix $\Phi$ of  \eqref{N-exponential} is a phase and defined up to a constant, as long as $N_1=\cos(2\Phi), N_2=\sin(2\Phi)$ are satisfied. On the other hand, the transformation matrix $\exp(\bar{\gamma}^0\Phi)=\cos(\Phi)+\bar{\gamma}^0\sin(\Phi)$ may take a minus sign according to the choice of $\Phi$, but the Lagrangian \eqref{lagrangian-hat} is independent of this choice. Moreover, the constant ambiguity does not change $\partial_0 \Phi $ and the minus signs in $\cos(\Phi)$, $\sin(\Phi)$ are compensated in the expressions of \eqref{M-hat}. As a result, this ambiguity has no effect on the Lagrangian or on the mass matrix.

Since $\hat{M}_2$ is antisymmetric, we can further perform an orthogonal transformation \cite{Nilles:2001fg,Nilles:2001ry} in order to eliminate this matrix
\begin{equation}
\hat{\Psi}=L\tilde{\Psi},\quad  \text{with } \left( \partial_0 + \hat{M}_2 \right)L=0\,.
\label{L-eq}\end{equation}
Thus, we arrive at a Lagrangian where the mixing comes only from the mass matrix, that is $\bar{\gamma}^0-$independent
\begin{equation}
    \mathcal{L}_{\tilde{\Psi}_1 \tilde{\Psi}_2}=-\bar{\tilde{\Psi}}\left[ \bar{\gamma}^{0} \partial_{0}+\mathrm{i} \bar{\gamma}^{i} k_{i}+L^{T} \hat{M}_{1} L\right]\tilde{\Psi}\,.
\end{equation}
$\tilde{M}\equiv L^{T} \hat{M}_{1} L$ is real and symmetric, hence it can be diagonalised by an orthogonal matrix $C$, with
\begin{equation}
    \mu=\operatorname{diag}(\mu_1,\mu_2)=C^T \tilde{M} C\label{mu}\,.
\end{equation}
The energy squared eigenvalues for the fermions are then of the form
\begin{equation}
E_i^2=k^2+\mu_i^2 \label{eigen-E}\,.
\end{equation}
Although the momentum squared is multiplied by 1, yet one cannot conclude from \eqref{eigen-E} that the sound speed of the physical fermions is the speed of light. 

To have a closer look at the propagation of physical degrees of freedom, we follow the approach in \cite{Nilles:2001fg,Nilles:2001ry,Ema:2016oxl} and expand $\tilde{\Psi}_i$ into creation and annihilation operators:
\begin{equation}
\tilde{\Psi}_i(x)=C_{i j} \int \frac{d^{3} \mathbf{x}}{(2 \pi)^{3 / 2}} e^{i \mathbf{k} \cdot \mathbf{x}}\left[U_{r}^{j \ell}(k, \eta) a_{r}^{\ell}(k)+V_{r}^{j \ell}(k, \eta) b_{r}^{\dagger \ell}(-k)\right]\,,
\label{expansion}
\end{equation}
where $r=\pm$ denotes the helicity components and a summation over repeated indices is understood.
The spinorial Fourier coefficients are written in terms of the helicity eigenfunctions $\psi_\pm$ and mode functions (matrices) $U_\pm$, $V_\pm$:
\begin{equation}
U_{r}^{i j} \equiv\left[\frac{U_{+}^{i j}}{\sqrt{2}} \psi_{r}, r \frac{U_{-}^{i j}}{\sqrt{2}} \psi_{r}\right]^{T}, \quad V_{r}^{i j} \equiv\left[\frac{V_{+}^{i j}}{\sqrt{2}} \psi_{r}, r \frac{V_{-}^{i j}}{\sqrt{2}} \psi_{r}\right]^{T}\,.
\end{equation}
Since $U_\pm$ and $V_\pm$ are related by charge conjugation invariance of $\tilde{\Psi}_i$, we can restrict ourselves to the mode equations of $U_\pm$. 
Taking the momentum along the $x^3-$axis and defining the antisymmetric matrix 
\begin{equation}
\Gamma \equiv C^{T} \partial_0{C}\,,
\label{defGamma}
\end{equation} 
the equations of motion of $\tilde{\Psi}_i$ result in
\begin{equation}
    \mathrm{i}\partial_0 \left(\begin{array}{c}
           U_+\\U_-
    \end{array} \right)= D \left(\begin{array}{c}
           U_+\\U_-
    \end{array} \right)\quad ,\quad D=\left(\begin{array}{cc}
     -\mathrm{i}\Gamma-\mu     &-k \mathbb{1}_2  \\
    -k   \mathbb{1}_2  &  -\mathrm{i}\Gamma+\mu  
    \end{array} \right) 
\end{equation}
where $D$ is a $4\times4$ hermitian matrix, whose diagonal blocks encode the time dependence. Its real eigenvalues are
\begin{equation}
    \begin{aligned}
&    \omega_{1,\pm}=\pm \left[\Gamma_{12}^2+k^2 +\frac{1}{2}\left(\mu_1^2+\mu_2^2 \right)+\left(\frac{1}{4}\left( \mu_1^2-\mu_2^2\right)^2+\Gamma_{12}^2 \left(4k^2+\left(\mu_1+\mu_2 \right)^2 \right) \right)^\frac{1}{2}\right]^\frac{1}{2}
\\   &\omega_{2,\pm}=\pm \left[\Gamma_{12}^2+k^2 +\frac{1}{2}\left(\mu_1^2+\mu_2^2 \right)-\left(\frac{1}{4}\left( \mu_1^2-\mu_2^2\right)^2+\Gamma_{12}^2 \left(4k^2+\left(\mu_1+\mu_2 \right)^2 \right) \right)^\frac{1}{2}\right]^\frac{1}{2}
    \end{aligned}\label{dispersion-general}
\end{equation}
where we denoted the $(1,2)$ element of $\Gamma$ by $\Gamma_{12}$. 

Note that when the fermions have degenerate mass $\mu_1=\mu_2=\bar{\mu}$, then from \eqref{mu}, $\hat{M}_1=\bar{\mu}\mathbb{1}_2$ and $\Gamma_{12}=0$. Thus, in this case we recover the relativistic dispersion relation with time-dependent mass:
\begin{equation}
    \omega_{1,\pm}= \omega_{2,\pm}=\pm \sqrt{k^2+\bar{\mu}^2}
\label{relat-dispersion}    
\end{equation}
In general, the dispersion relations of the physical fermions in \eqref{dispersion-general} are very different from that in \cite{Kolb:2021nob}, describing the vacuum helicity-$1/2$ mode of the gravitino. We recall the latter for clarity:
\begin{equation}
\omega_{k} \equiv \sqrt{c_{s}^{2} k^{2}+a^{2} m_{3/2}^{2}}\label{kolb-dispersion}
\end{equation}
One of the arguments for catastrophic gravitino production at $c_s=0$ is based on the adiabaticity violation. The dimensionless coefficient of non-adiabaticity is defined as  \cite{Kofman:1997yn}
\begin{equation}
    \mathcal{A}_k\equiv \frac{\partial_0 \omega_k}{\omega_k^2}
\label{non-adia}\end{equation}
Indeed, in \eqref{kolb-dispersion}, the momentum is multiplied by $c_s$, so when $c_s=0$, the coefficient of non-adiabaticity is independent of $k$, and can even exceed one under some circumstances implying particle production with an arbitrarily large momentum. This is, however, not the case here: the sound speed of \eqref{kolb-speed} does not enter explicitly\footnote{$\Gamma_{12}$ can potentially depend on $c_s$, because $\Gamma$ is related to the diagonalization of the mixing matrices, which contain $\Delta$.} the physical dispersion relations, and cannot suppress the momentum dependence when it vanishes. Therefore, $c_s$ is not, at least not directly, responsible for the divergent particle production.

To see if particles of arbitrarily large momenta can actually be produced, we assume that the only source of non-adiabaticity is due to the variations of frequencies of the two physical fermions, with the coefficient $\mathcal{A}_k$  being a sum of them. We then consider the limit of high momenta $k\gg\mu_i$ and $k\gg \Gamma_{12}$, where $\Gamma_{12}$ is roughly the time derivative of the logarithm of masses (see its definition \eqref{defGamma}), defining a scale related to the variation of masses.
In this limit, the non-adiabaticity coefficient at leading order becomes
\begin{equation}
    \mathcal{A}_k\equiv \frac{\partial_0 \omega_{1,+}}{\omega_{1,+}^2}+\frac{\partial_0 \omega_{2,+}}{\omega_{2,+}^2}\approx -6 \Gamma_{12}\partial_0 \Gamma_{12}\frac{1}{k^3}
\end{equation}
implying that $\mathcal{A}_k$ falls as $k^{-3}$, and thus particles with arbitrarily large $k$ cannot be produced.

\section{Examples}
\label{sec:examples}

\subsection{Two chiral multiplets}

Given two chiral multiplets $(\chi_1, \phi_1)$, $(\chi_2, \phi_2)$, we investigate the case with $\Delta=1$ \textit{at all times}, so that the sound speed defined in  \eqref{kolb-speed} vanishes, and according to \cite{Kolb:2021nob,Kolb:2021xfn}, the gravitino production is expected to diverge. In this toy example, the physical fermions as well as their equations of motion can be explicitly obtained. For simplicity,  the K\"ahler potential is taken to be canonical, so that the expression of $\Delta$ becomes:
\begin{equation}
        \Delta=\frac{2}{\alpha}\left( m_1\dot{\phi}_2-m_2\dot{\phi}_1\right) =1
\end{equation}
On the other hand, from \eqref{notations} on sees that $\Delta=1$ is equivalent to the conditions $\alpha_1=\alpha_2=0$. The latter implies that the gravitino mass is constant which is equivalent to the condition 
$m_1\dot\phi_1+m_2\dot\phi_2=0$. The former implies $\dot\phi_1^2+\dot\phi_2^2=m_1^2+m_2^2$, where we used the expression \eqref{potential} for the potential in the absence of $D$-term contribution.
One possible setup for a solution is therefore $m_2=\dot{\phi}_1=0$ and  $m_1= \dot{\phi}_2\neq0$, in other words, $\phi_1$ breaks supersymmetry via its $F$-term and $\phi_2$ via its kinetic term, by the same amount. 

The mixing matrix $N$ then takes the form
\begin{equation}
    N=\bar{\gamma}^0\left(\begin{array}{cc}
        0 & -1 \\
        -1 & 0
    \end{array} \right)\,,
\label{N-twochiral}
\end{equation}
while, for convenience, we write $M$ as
\begin{equation}
    M=\mathbb{1}_4\left(\begin{array}{cc}
        M_{11} & 0 \\
        0 & M_{22}
    \end{array} \right)\quad,\quad M_{11},M_{22}\neq0\,.
\end{equation}
The angle $\Phi$ in \eqref{N-exponential} is constant because $N$ does not depend on time; we then choose
\begin{equation}
    \Phi =\frac{3\pi}{4} \left( \begin{array}{cc}
    0 &1  \\
1 &0 
\end{array}\right),\quad \operatorname{exp}(\bar{\gamma}^0\Phi )=-\frac{1}{\sqrt{2}}\mathbb{1}_2 + \frac{1}{\sqrt{2}} \bar{\gamma}^0 \left( \begin{array}{cc}
     0&1  \\
     1&0 
\end{array}\right)\,.
\end{equation}
Upon the unitary transformation $\hat{\Psi}=\operatorname{exp}(\bar{\gamma}^0\Phi) \Psi$, we obtain the new mass matrix $\hat{M}$ with
\begin{equation}
    \hat{M}_1=\frac{M_{11}+M_{22}}{2}\mathbb{1}_2
    \quad ,\quad \hat{M}_2=\frac{M_{11}-M_{22}}{2}\left( \begin{array}{cc}
         0& 1 \\
         -1&0 
    \end{array}\right)\,.
    \end{equation}

We now look for the orthogonal matrix $L$ cancelling the $\bar{\gamma}^0$ component of $\hat{M}$. Parametrising $L$ by an angle $\tau$, from \eqref{L-eq} we get
\begin{equation}
    L=\left(\begin{array}{cc}
        \operatorname{cos}\tau(\eta) &- \operatorname{sin}\tau (\eta)\\
         \operatorname{sin}\tau(\eta)& \operatorname{cos}\tau(\eta)
    \end{array} \right),\quad \tau^\prime(\eta)=\frac{M_{11}-M_{22}}{2}
\end{equation}
Once such an orthogonal transformation is found,  the Lagrangian in the $\tilde{\Psi}=L^T \operatorname{exp}(\bar{\gamma}^0\Phi)\Psi$ basis becomes
\begin{equation}
    \mathcal{L}_{\tilde{\Psi}_1\tilde{\Psi}_2}=-\bar{\tilde{\Psi}}\left[\bar{\gamma}^{0} \partial_{0}+i \bar{\gamma}^{i} k_{i}+  \frac{M_{11}+M_{22}}{2}   \right]\tilde{\Psi}\,.
\end{equation}
Thus, the mass matrix in this particular case is diagonal and one concludes that $\tilde{\Psi}=\{\tilde{\Psi}_1,\tilde{\Psi}_2\}$ are the physical fermions. They have degenerate mass and their equations of motion are decoupled:
\begin{equation}
    \left[\bar{\gamma}^{0} \partial_{0}+i \bar{\gamma}^{i} k_{i}+  \frac{M_{11}+M_{22}}{2}   \right]\tilde{\Psi}_j=0,\quad j\in\{1,2\}
\label{eom-psi-tilde}\end{equation}
These are just standard Dirac equations with time-dependent mass, similar to the transverse mode of the gravitino. As a result, the physical fermions, which are linear combinations of $\theta$ and $\Upsilon$, have the dispersion relation
\begin{equation}
    \omega^2=k^2+\left(  \frac{M_{11}+M_{22}}{2} \right)^2\,.
\end{equation}
The coefficient of non-adiabaticity is suppressed by large momenta, and particle production is not expected to be divergent in this case, despite the fact that the speed of sound \eqref{Sound-speed-1} associated to $\theta$ vanishes.

\subsection{One chiral multiplet and one vector multiplet}

Another example with two fermions is one chiral multiplet $(\chi_1,\phi_1)$ accompanied by a vector multiplet with non-vanishing $D$-term. The simplest model one may consider is that of a constant (Fayet-Iliopoulos) $D$-term when the vector multiplet gauges the R-symmetry~\cite{Freedman}. One can then consider two possibilities. The first consists of a neutral chiral multiplet, in which case the superpotential vanishes. Again, we take the K\"ahler potential to be canonical. The scalar potential is then
\begin{equation}
V =V_{D}=\frac{1}{2}g^2 \mathcal{P}^2
\end{equation}
with $\mathcal{P}$ constant.
Since the gravitino is massless in this model, all mass terms appearing in \eqref{F-Dterm} vanish, and we have $\hat{F}=0$. Let
\begin{equation}
    w\equiv \frac{p}{\rho}, \quad \text{with}\quad  p=|\dot{\phi}|^2-V_D,\quad \rho=|\dot{\phi}|^2+V_D \,.
\end{equation}

In this model, the scalar equation of motion can be easily solved. We have
\begin{equation}
    \ddot{\phi}+3H\dot{\phi}=0
\end{equation}
leading to $\dot{\phi}=e^{-3Ht}$, where the constant of integration is absorbed by a redefinition of the origin of time. The expression of $w$ is then
\begin{equation}
    w(t)=\frac{2e^{-6Ht}-g^2\mathcal{P}^2}{2e^{-6Ht}+g^2\mathcal{P}^2}\label{wt}\,.
\end{equation}
Moreover, the parameters in \eqref{notations} and $\Delta$ can be written as
\begin{equation}
    \begin{aligned}
\hat{A}=\hat{A}_1=w,\quad \hat{B}=\hat{B}_1=-\frac{3\dot{a}}{2}\left(1-w^2 \right),\quad\Delta^2=1-w^2\,.
    \end{aligned}
\end{equation}
The sound speed defined in \eqref{kolb-speed} is therefore simply given by $w$, and the mixing matrices in this case are
\begin{equation}M=0,\quad N=\left(\begin{array}{cc}
w   &0  \\
     0& w
\end{array}\right)+\bar{\gamma}^0\left(\begin{array}{cc}
    0 & -\sqrt{1-w^2}  \\
     -\sqrt{1-w^2}&0 
\end{array}\right)\label{MN-Dterm}
\end{equation}
It follows that the physical fermions production and their equations of motion are determined only by $w$. Here, we study some particular limits:
\begin{itemize}
\item
When $w\rightarrow 1$, the pressure and the energy density are equal, implying that the $D$-term is vanishing. This limit amounts to a theory with a single chiral multiplet.  $N$ is the identity matrix whereas $M=0$, so $\Psi_i$ are the physical fermions described by the massless Dirac equation and propagating at the speed of light.\footnote{This agrees also with the literature for the one chiral multiplet case; see the discussion in Section \ref{sec:dispersion}.}
\item
On the other hand, the same situation occurs for $w\rightarrow-1$, which means that $|\dot{\phi}|^2\rightarrow0$, or equivalently $t\rightarrow +\infty$ in \eqref{wt}, so that maximal symmetry is unbroken.
The equations of motion of the physical fermions differ from the massless Dirac equation by a minus sign in front of $\bar{\gamma}^ik_i$, but the dispersion relation  $\omega^2=k^2$ is unchanged compared to the previous case.
\item
Another special value is $w=0$, corresponding to zero pressure and $e^{-6Ht}=V_D$, which is always satisfied for a certain time $t$. The mixing matrix $N$ is identical to  \eqref{N-twochiral} for two chiral multiplets. The diagonalisation can be carried out in exactly the same way and we obtain two physical fermions with degenerate mass (massless here). Their equations of motion are the massless Dirac equation  and we do not expect a divergent particle production, even though $c_s=0$.
\end{itemize}
More generally, the matrix $\Phi$ in \eqref{N-exponential} can be chosen as
\begin{equation}
    \Phi=-\frac{1}{2}\operatorname{arccos}(w)\left( \begin{array}{cc}
       0  &1  \\
         1& 0
    \end{array}\right)
\end{equation}
while for $w\neq \pm 1$
\begin{equation}\hat{M}=\partial_0 \Phi =\frac{\partial_0 w}{2\sqrt{1-w^{2}}}\left(
\begin{array}{cc}
     0&1  \\
     1&0 
\end{array}\right)\,.
\end{equation}
The matrix $\hat{M}$ has no $\bar{\gamma}^0$ component and can be diagonalised by a constant orthogonal matrix $C$. Thus $\Gamma=0$, leading again to relativistic dispersion relations.
The details of particle production can be worked out by doing the expansion \eqref{expansion}, and solving numerically differential equations for the Bogolyubov coefficients, which will not be discussed here.

Finally, we comment briefly on the second possibility where the chiral field has a non-vanishing R-charge and the K\"ahler potential is non-canonical. Consider for instance a realistic model of inflation driven by supersymmetry breaking \cite{Antoniadis:2017gjr}, where the K\"ahler potential and the superpotential are 
\begin{equation}
    \begin{aligned}
    K=\phi^1\phi_1+A (\phi^1\phi_1)^2,\quad W=f \phi_1
    \end{aligned}
\label{KW}
\end{equation}
with $f$ a constant and $|\phi_1|$ playing the role of the inflaton, while its phase is absorbed in the gauge field to make it massive. 
$A$ is a small positive constant, so that the potential has a maximum at the origin allowing hilltop inflation with the slow-roll parameter $\eta$ controlled by $A$.
In this case, the $D$-term part of the potential is given by
\begin{equation}
    V_D=\frac{q^2}{2}\left(1+\phi^1 \phi_1 + 2A (\phi^1 \phi_1)^2\right)^2\,,
\end{equation}
with $q$ a constant parameter corresponding to the R-charge of $\phi_1$, that must be small compared to the $F$-term so that $V_D$ is subdominant during inflation.
It follows that $\Delta$ has the form:
\begin{equation}
 \begin{aligned}
 \alpha^{2} \Delta^{2}=4V_D (1+4A\phi^1\phi_1) \dot{\phi}^1 \dot{\phi}_1\quad ,\quad \alpha^2=\left( \rho+3 M_{P}^{-2}|m|^{2}\right)^2\,.
 \end{aligned}
\end{equation}

Note that it appears possible to have $\Delta^2<0$ for some negative values of $A$, leading to $c_s>1$ according to equation \eqref{kolb-speed}. However, this region is unphysical since the K\"ahler metric becomes negative. This is actually similar to the situation that can be obtained in pathological models where $\Upsilon$ is dropped out by constraints, leading to $c_s>1$ \cite{Dudas:2021njv}.

We will now investigate again the case of $\Delta=1$ at all times. From \eqref{notations}, one sees that the condition $\alpha_2=0$ implies that the gravitino mass is constant with $m_1=0$, which is equivalent that $\phi_1$ has vanishing $F$-term and breaks supersymmetry only by its kinetic energy. This may indeed be satisfied around the vacuum at the minimum of the potential after the end of inflation. There, $\phi_1$  is in general far from the maximum at the origin where corrections to the K\"ahler potential \eqref{KW} become important and change its form. On the other hand, the condition $\alpha_1=0$ implies $|\dot{\phi}_1|^2=V_D$ where the latter is now given by $V_D=(q^2/2)(1+K^1\phi_1)^2$ with $K^1\equiv\partial K/\partial\phi_1$. The analysis of the two fermions $\theta$ and $\Upsilon$ can now proceed as in the previous subsection of two chiral multiplets, giving rise to two decoupled equations of motion with a relativistic dispersion relation.

\section{Conclusions}
\label{Conclusions}

We have studied the equations of motion for the longitudinal modes of gravitinos in  supergravity models where supersymmetry is linearly realised but spontaneously broken. We have considered the general case of two supermultiplets. One contains a scalar field $\phi$ that has non-vanishing kinetic energy, $\partial_\mu \phi \neq 0$. In a cosmological background, this scalar can be identified with the inflaton which is time dependent. The other multiplet is at the origin of the gravitino mass in the vacuum at late times. We have found that, after diagonalisation of the Hamiltonian, in all cases the dispersion relations of the propagating fermions take relativistic forms with in general a time-dependent mixing mass matrix. While this might be expected, it is shown here explicitly. Such cases are not expected to show a catastrophic production of gravitinos.

We did not discuss here the non-linear models as those considered in \cite{Ferrara:2015tyn,Dudas:2021njv,Terada:2021rtp} as it is not clear to us which microscopic supergravity Lagrangian is at the origin of the constraint imposed there on the inflaton superfield. We note that in these cases one ends up with one fermion propagating in peculiar backgrounds. 
The result of \cite{Kolb:2021xfn,Kolb:2021nob} constrains the background on which Rarita-Schwinger fields are allowed to propagate.

\section*{Acknowledgements}
Work partially performed by I.A. as International Professor of the Francqui Foundation, Belgium. The work of K.B. is supported by the Agence Nationale de Recherche under grant ANR-15-CE31-0002 ``HiggsAutomator''.
I.A. would like to thank Toine Van Proeyen for enlightening discussions.

\appendix

\section{Notations}
\label{appendix}

Our notations follow mostly that in \cite{Kallosh:2000ve}, where the flat space $\gamma$-matrices are 
\begin{equation}
\bar{\gamma}^{0}=\left(\begin{array}{cc}
\mathrm{i} \mathbb{1}_2 & 0 \\
0 & -\mathrm{i} \mathbb{1}_2
\end{array}\right), \quad \bar{\gamma}^{i}=\left(\begin{array}{cc}
0 & -\mathrm{i} \sigma_{i} \\
\mathrm{i} \sigma_{i} & 0
\end{array}\right),\quad {\gamma}_{5}=\left(\begin{array}{cc}
0 & -\mathbb{1}_{2} \\
-\mathbb{1}_{2} & 0
\end{array}\right)
\end{equation}
The Minkowski metric has signature $(-,+,+,+)$, and for cosmological applications, we used the FLRW metric. The curved space $\gamma$-matrices, noted $\gamma^\mu$, are then related to the flat space $\gamma$-matrices by $\gamma^{\mu}=a^{-1} \bar{\gamma}^{\mu}$.
The left and right projections are defined as
\begin{equation}
P_{L}=\frac{1}{2}\left(1+\gamma_{5}\right)\quad , \quad P_{R}=\frac{1}{2}\left(1-\gamma_{5}\right)    
\end{equation}
Note that for chiral fermions, $P_L\chi_i=\chi_i$ and $P_R\chi^i=\chi^i$. The charge conjugation matrix in this convention is given by $C=\bar{\gamma}^0\bar{\gamma}^2$. Some useful charge conjugates are
\begin{equation}
    \begin{aligned}
&    \chi_{i}^{C}=\chi^{i}, \quad \phi_{i}^{C}=\phi^{i},\quad \mathcal{P}^C=\mathcal{P},\quad \lambda^C=\lambda\\
 &\bar{\gamma}_{\mu}^{C}=\bar{\gamma}_{\mu}, \quad \gamma_{5}^{C}=-\gamma_{5}, \quad P_{L}^{C}=P_{R}   \end{aligned}
\end{equation}

In the Lagrangian \eqref{lagrangian}, the covariant derivative of the scalar is  $\hat{\partial}_{0}=\partial_{0}-\frac{i}{2} A_{0}^{B} \gamma_{5}$. In the cosmological context considered, the spatial derivatives of the scalar vanish, and for real backgrounds we have $A_{0}^{B} =0$ and thus $\hat{\partial}_{0}={\partial}_{0}$.
Keeping the above simplifications, the covariant derivatives acting on the chiral fermions, the gaugino and the gravitino are respectively
\begin{equation}
\begin{aligned}&D_{\mu} \chi_{i} \equiv\left(\partial_{\mu}+\frac{1}{4} \omega_{\mu}^{ab} \bar{\gamma}_{ab}\right) \chi_{i}+\Gamma_{i}^{j k} \chi_{j} \partial_{\mu} \phi_{k}
,\quad 
\mathcal{D}_{\mu} \lambda=\left(\partial_{\mu}+\frac{1}{4} \omega_{\mu}^{a b} \bar{\gamma}_{a b}\right) \lambda  \\&
D_{\mu} \psi_{\nu}=\left(\left(\partial_{\mu}+\frac{1}{4} \omega_{\mu}^{ab} \bar{\gamma}_{ab}+\frac{1}{2} \mathrm{i} \gamma_{5} A_{\mu}\right) \delta_{\nu}^{\lambda}-\Gamma_{\mu \nu}^{\lambda}\right) \psi_{\lambda}
\end{aligned}\end{equation}
where
$\omega^{ab}_\mu$ stands for the spin connection and $A_\mu$ is the $U(1)$ gauge field. The Christoffel connection   $\Gamma^\lambda_{\mu\nu}$  differs from the Kähler connection, where the latter corresponds to $\Gamma_{i}^{j k} \equiv g^{-1}{ }_{i}^{l} \partial^{j} g_{l}^{k}$. We use the notation $\bar{\gamma}_{ab}\equiv \left[\bar{\gamma}_{a}, \bar{\gamma}_{b}\right] / 2$.

Having introduced the Kähler covariant derivative $\mathcal{D}^i$, the mass terms are
\begin{align}&m^{i} \equiv \mathcal{D}^{i} m=\partial^{i} m+\frac{\partial^{i} {K}}{2 M_{\mathrm{P}}^{2}} m ,\quad m^{i j} \equiv \mathcal{D}^{i}\mathcal{D}^{j} m=\left(\partial^{i}+\frac{ \partial^{j} K}{2 M_{\mathrm{P}}^{2}}\right) m^{j}-\Gamma_{k}^{i j} m^{k}\\&m_{i \alpha}=-\mathrm{i}\left[\partial_{i} \mathcal{P}-\frac{1}{4}(\operatorname{Re} f)^{-1 } \mathcal{P} f_{i}\right],\quad m_{R, \alpha \beta}=-\frac{1}{4} f_{ i} g_{j}^{-1 i} m^{j}\end{align}
where the subscript $i$ in $f$ denotes derivative with respect to $\phi^i$.



\end{document}